\documentclass{ws-p8-50x6-00} 
\usepackage{graphicx}

\begin{document} 

\title{Chaotic Phenomena in Astrophysics and Cosmology}

\author{V.G.Gurzadyan} 

\address{ICRA, Department of Physics, University of Rome "La Sapienza",
00185 Rome, Italy 
and 
Department of Theoretical Physics
Yerevan Physics Institute, 375036 Yerevan,
Armenia
}

\maketitle 


{\it Lectures at the Xth Brazilian School of Cosmology and Gravitation, July-August, 2002; Published in "Cosmology and Gravitation", Eds.M.Novello, S.E.P.Bergliaffa, pp.108-124, AIP, New York, 2003.}

\section{Introduction}

Chaos is a typical property of many-dimensional nonlinear systems. Its role it 
revealed in various problems of astrophysics and cosmology. 
Chaos made to revise the two-hundred year old views on the evolution of Solar system. Theory of interstellar matter, dynamics of star clusters and galaxies at present cannot be considered without chaotic effects.

Astronomical topics themselves had remarkable impact on the development of chaotic dynamics.
The Henon-Heiles system, one of the first systems with revealed chaotic properties,
was proposed for the study the motion of a star in a galactic potential. 
Much earlier, Poincare's classical
work on the foundations of the theory of dynamical systems emerged from the
problem of small perturbations in the planetary dynamics.

In the present lectures I will discuss only several astrophysical and cosmological problems.
The choice of the problems is determined with the aim, first, to cover as broad topics as possible,
second, to show the diversity of approaches and mathematical tools. I will 
start from planetary dynamics, moving to galactic dynamics, to cosmology,
and to the instability in the
Wheeler-DeWitt superspace. For pedagogical reasons I will describe the 
techniques, such as the estimation Kolmogorov-Sinai
entropy, the dealing with hyperbolicity in pseudo-Riemannian spaces, so that 
they can applied for any other problems. Obviously, numerous other problems, 
methods and results remain out of these lectures, most of them, however, can be
traced from the references; for chaos see\cite{sagdeev,lich,ott,zaslav}, for
applications of our interest see\cite{GPfen,hobill,book,ChaoticUniv}.

I will start from a brief review of the elements of theory of dynamical systems, 
to introduce the main concepts used in the subsequent chapters.

\section{Elements of Ergodic Theory}
 
\subsection{Dynamical systems}

Ergodic theory is the metric theory of dynamical systems, i.e.
which deals with spaces for which a measure is defined but
not a metric. 

In the following brief account of elements of  
smooth ergodic  theory, we will concentrate on the
classification  of  dynamical  systems  by the degree of their
statistical properties; for details see\cite{arn,ds,katok}.
 
The key concept is obviously, that of the dynamical system. Initially the dynamical systems were understood as mechanical systems, 
however later that term was generalized to variety of physical systems of non-mechanical origin.
Cosmological solutions of Einstein equations can be considered as such examples.

Dynamical system $(M, {\cal B}, \mu, T)$ is considered defined 
if $M$ is a smooth manifold, $\cal B$ is a $\sigma$--algebra of measurable sets on $M$, and $\mu$ is
a complete measure on $\cal B$, and $T^t$ is a
one-parameter group of diffeomorphisms
defined by vector field {\bf v}
\begin{equation}
  {\bf v}(x)=\frac{dT^tx}{dt}.
\end{equation}
One-parameter groups are called flows, by the term borrowed from hydrodynamics, 
and below we will give elements of the classification of flows. 
The apparent abstractness of the definition implies
quite general and natural properties for physical systems. 
   
\subsection{Classification of dynamical systems, mixing, relaxation}

Flows are called {\it ergodic}, if for any measurable invariant set $A$
\begin{equation}
T^t A=A=T^{-t}A,
\end{equation}
its measure $\mu$ takes only the values
\begin{equation}
\mu(A)=\left\{ \begin{array}{ll}
               0 \\ 1
               \end{array}\ .
       \right.
\end{equation}

One can show that for measure-preserving  ergodic flows the
time-average almost everywhere equals the phase space average
\begin{equation}
\int_M f d\mu = \lim_{t \rightarrow \infty} \frac{1}{t} \int_{0}^{t}
f(T^{-\tau} (x)) d\tau.
\end{equation}
In physical literature this property is often considered as a
definition of an ergodic system since it is enough and sufficient.
The property of ergodicity is one of rare definitions of smooth
ergodic theory which can be generalized also for spaces with
infinite measure.
Ergodicity, however, is a weak statistical property and therefore is
less important for actual physical problems.
 
The far more importance for statistical physics of another property, mixing,
has been established firstly by Gibbs.

Ergodic theory provides definitions for mixing of various degrees.

{\it  Weak
mixing} is indicated by the condition for $\forall f,g\in  L^2$
\begin{equation}
   \lim_{t \rightarrow \infty} \frac{1}{t} \int_{0}^{t}
         \left[ \int_{M}f(T^{-\tau}x) g d\mu
         - \int_{M}f d\mu \int_{M} g d\mu\right]^2 d\tau=0.
\end{equation}

The 'weakness' of the property of weak mixing can be seen from the following limit
\begin{equation}
\lim_{t \rightarrow \infty} \frac{1}{t}
\int_{0}^{t} \mid
\mu(T^{-\tau} A \cap B) - \mu (A) \mu (B) \mid d\tau = 0,
\end{equation}
implying that $T^tA$ becomes independent of the set $B$ only if some parts
of the trajectory are not taken into account.

Note the absence of the factor '1/t' and hence increase in the
convergence rate in the definition
of the property of {\it mixing} 
\begin{equation}
\lim_{t \rightarrow \infty}
\int_{M}f(T^t x) g d\mu = \int_{M}f d\mu \int_{M} g d\mu.
\end{equation}
Analogically the property of {\it m-fold mixing} for $m$ functions is 
generalized as follows
\begin{equation}
\lim_{t_1,...,t_m \rightarrow \infty}
\int_{M}f_0 f_1(T^{t_1}x)...f_m (T^{t_1 +...+t_m} x) d\mu =
\prod_{i=0}^{m} \int_{M}f_i d\mu .
\end{equation}
These properties  describe  systems  with  increasing
statistical properties in the sense that, systems with mixing  possess
the property of weak mixing, and those with $n$-fold mixing also that of mixing
and weak mixing but not vice versa.
 
Systems with mixing  are evidently also  ergodic  ones.  However,
for  the   systems   with   mixing, as opposed to ergodic  ones,
a set $A\in \cal B$ evolves in such a way
(preserving its measure and connection)
that the measure of the part  which intersects the set $B\in \cal B$
tends in time to be proportional to the measure of $B$
\begin{equation}
\lim_{t \rightarrow \infty} \frac{\mu (T^t A \bigcap B)}{\mu(A)} =
\mu(B).
\label{mix}
\end{equation}
 
Compare this limit with the following one for ergodic systems
\begin{equation}
\lim_{t \rightarrow \infty} \frac{1}{t}\int_{0}^{t}
\mu(T^{-\tau} A \cap B) d\tau = \mu(A)\mu(B),
\end{equation}
The latter limit is said to converge {\it in the Cesaro
sense}, while the limit for the mixing case is ordinary converging.
In other words, for ergodic
systems the  initial fluctuations tend  to zero
only in the time-average,
for mixing systems their absolute value decreases as well.

Hence the property of mixing guarantees the existence of a final state of measure $\mu$ to which
\begin{equation}
\mu_t(A) = \mu_0(T^tA), \ A\in \cal B     
\end{equation}
tends smoothly, so that
\begin{equation}
\lim_{t \rightarrow \infty}  \int_{M}
f d\mu_t =  \int_{M}f d\mu .
\end{equation}
In physical terminology the final  state is called $\mu$
{\it equilibrium}, the  process  of  tending  to  that   state  is the {\it relaxation}.
   
Even stronger statistical property, K-mixing, possess K-systems (Kolmogorov systems) systems for which the following limit
\begin{equation}
\lim_{t \rightarrow \infty} \sup  \mid
\mu( A \cap B) - \mu (A) \mu (B) \mid =0
\end{equation}
exists, where the upper limit is taken for the smallest
$\sigma$-algebra containing $T^{t_0}A$ for $t < t_0$.
Kolmogorov systems possess $n$-fold mixing of arbitrary $n$.

Strongest statistical  properties  are possessed by hyperbolic  (Axiom-A),
Anosov, and Bernoulli systems.\footnote{As mentions Smale,\cite{Smale} he had found the 
horseshoe transformation, the classic example of Axiom-A systems in Brazil, on Rio beaches.}

We will define Anosov systems\cite{an} which by their statistical properties
are equivalent (isomorphic) to Bernoulli shifts. 
  
The flow $f^t$ is of  Anosov  type  if for its
all trajectories $\{f^t\}$ there  exist
subspaces $E^s(f^t(x)), E^u(f^t(x))$ of the tangential space  $TM_{f^t(x)}$,
and numbers
$C  > 0, \lambda > 0$, such that
\begin{eqnarray*}
& & TM_{f^t(x)} = E^s(f^t(x)) \oplus E^u(f^t(x)) ,\\ 
& & df^{\tau} E^s(f^t(x)) = E^s (f^{t+\tau}(x)), \qquad
df^{\tau} E^u(f^t(x)) = E^u (f^{t+\tau}(x)),
\end{eqnarray*}
and for all $t > 0$, one has
\begin{eqnarray*}
& &
\parallel df^{t} v \parallel \leq C e^{-\lambda t} \parallel v \parallel,
\qquad v \in E^s;\\
& &
\parallel df^{t} v \parallel \geq C^{-1} e^{\lambda t}
\parallel v \parallel,
\qquad v \in E^u.
\end{eqnarray*}
The subspaces $E^s$ and $E^u$ are stable (converging) and unstable
(expanding) subspaces.

For physical systems this definition implies exponential instability
at each point of the phase trajectory and at any small perturbation.
Anosov systems are subclass of hyperbolic systems
and they possess an important property of structural stability.
Roughly it means that the perturbed systems possess the property
of the unperturbed one, i.e. the strong instability acts towards
preserving of the properties of the system. Though strictly speaking 
the conditions
of Anosov systems are never or almost never are satisfied for
real physical systems, nevertheless it appears that the structural
stability can be peculiar to certain types of strongly instable physical systems.

Geodesic flow on a compact  manifold  with  negative  constant  curvature
is  an  example  of an Anosov system, and was studied long ago by Hadamard, Hopf  and  Hedlund. Their works had inspired Krylov\cite{Krylov} to apply those ideas to physical systems.

If  the  systems  with  mixing  can  tend  to
equilibrium by any law (e.g. polynomial), 
hyperbolic systems tend to that state exponentially. 
 
\subsection{Kolmogorov-Sinai entropy}

The problem of distinguishing different features of
dynamical  systems,  and the
formulation of corresponding characterizing criteria
is a  central  one in
ergodic theory. Much efforts in this direction were concentrated on
the study of spectral  properties  of  dynamical
systems, until in 1958,  Kolmogorov  discovered  the  new  metric
invariant, the entropy.
 
Consider the entropy of a splitting $\xi_i $ of the  measurable  manifold M
\begin{equation}
H(\xi)= \sum_{i=1}^{d} \mu(\xi_i) \ln (\xi_i),
\end{equation}
where $\xi_i\in{\cal B}$ and
\begin{equation}
 \xi_i\bigcap\xi_j = \emptyset\ \mbox{ if } i\not=j,
\end{equation}
\begin{equation}
 \bigcup_{i=1}^{d}\xi_i = M.
\end{equation}
Then the Kolmogorov-Sinai (KS) entropy $h$ is the limit
\begin{equation}
h(f)=\sup \lim_{n \rightarrow \infty} \frac {1} {n} H(\xi^n),
\end{equation}
where
\begin{equation}
   \xi^n=\bigvee_{j=0}^{n-1}f^{-j}\xi,
\end{equation}
and the upper limit is taken over all measurable splittings.
 
Dynamical systems with positive KS-entropy $h>0$ are usually called {\it
chaotic}, while those with $h=0$  are called {\it  regular}  ones.
In particular, Anosov and Kolmogorov systems, which are typical
systems with mixing, have positive KS-entropy $h>0$,
while most of only ergodic ones have $h=0$.
Therefore the latter are not considered to be chaotic
according to this definition.
 
For the above mentioned geodesic flows on spaces with constant negative
curvature $R<0$, the KS-entropy equals
\begin{equation}
h = \sqrt{-R}.
\end{equation}
 
The KS-entropy is related to the Lyapunov characteristic exponents $\lambda_i$
via the Pesin formula
\begin{equation}
h(f)=\int_{M} \sum_{\lambda_i(x)>0} \lambda_i(x) d\mu(x);
\end{equation}
we see that a system with at least  one  non-zero Lyapunov exponent
has positive KS-entropy. The use of Lyapunov exponents for many-dimensional systems
is not always well defined, nevertheless it was efficiently applied for stellar
systems\cite{Pfennig}.
 
Finally let us mention another important characteristic of  dynamical
systems, the correlation function, defined by
\begin{equation}
b_{g,g^{'}}(t)=\int_{M} g(f^tx)g^{'}(x)d\mu-
\int_{M}g(x)g^{'}(x)d\mu.
\end{equation}
Although  at  present
estimates of the correlation functions (including numerical results on
some   billiards) exist only  for  a few  dynamical  systems,
for   Anosov   systems   it  has   been shown    that
the correlation   functions    decay    exponentially,    i.e.,    $\exists
\alpha_{g,g^{'}}, \beta, t>0 $ so that
\begin{equation}
   |b_{g,g^{'}}(t)|\leq \alpha_{g,g^{'}}\exp(-\beta t),
\end{equation}
where
\begin{equation}
\beta \simeq h(f).
\end{equation}

\section{Chaotic Solar System}

Results obtained in recent decades have revealed the crucial role of chaotic effects
in planetary dynamics. 
For detailed reviews I would refer to \cite{murray,lecar,morb}, where various evidences
of chaos, particularly in the asteroid belt, in the motion of comets, 
are discussed, along with the methods of overlapping resonances and estimation
of Lyapunov exponents, Wisdom-Holman symplectic mapping and other techniques
used in those studies.

Before considering the stability of the Solar system, let us formulate the
two key theorems, Poincare's and Kolmogorov's, which were crucial in the
efforts on this long-standing problem.

N-dimensional system is considered as integrable
if its first integrals $I_1,...,I_N$ in involution are known,
i.e. their Poisson brackets are zero. As follows from the Liouville theorem if the set of 
levels 
$$
M_I=\{I_j(x)=I^0_j, j=1,...,N\}
$$
is compact and connected, then it is diffeomorphic to N-dimensional torus
$$
T^N=\{(\theta_1,...,\theta_N), modd 2\pi\}, 
$$
and the Hamiltonian system performs a conditional-periodic motion on $M_I$.
Poincare theorem  states that {\it for a system with perturbed Hamiltonian
\begin{equation}
H(I, \varphi, \epsilon) = H_0(I) + \epsilon H_1(I, \varphi, \epsilon),
\end{equation}
where $I, \varphi$ are action-angle coordinates, at small $\epsilon > 0$ 
no other integral exists besides the one of energy $H=const$, if
$H_0$ fulfills the nondegeneracy condition,
\begin{equation}
det |\partial \omega /\partial I|\not= 0,
\end{equation}
i.e. the functional independence of the
frequencies $\omega=\partial H_0/\partial I$ of the torus over which the conditional-periodic winding is performed.}

Though this theorem does not specify the behavior of the trajectories of the system
on the energy hypersurface, up to 1950s it was widely believed that
such perturbed systems have to be chaotic. 

Kolmogorov's theorem\cite{kolm} of 1954, the main theorem of Kolmogorov-Arnold-Moser theory, showed
that at certain conditions the perturbed Hamiltonian systems can remain stable. 

It states:

{\it If the system (24) satisfies the nondegeneracy condition (25) and
$H_1$ is an analytic function, then 
at enough small $\epsilon > 0$ most of non-resonant tori, i.e.
tori with rationally independent frequencies satisfying the condition
\begin{equation}
\sum n\omega_k\not=0,
\end{equation}
do not disappear and the measure of the complement
of their union set $\mu(M)\rightarrow 0$ at $\epsilon\rightarrow 0$.}

KAM-theory was initially considered as supporting the views on the stability of the Solar system, though
it says nothing about the limiting value of the perturbation $\epsilon$.

However, though the level of direct applicability of the KAM theory for the Solar system remains not clear,
it appears that, the joint application both of theoretical and numerical methods at present computer's possibilities
is rather efficient. 

The frequency map technique developed by Laskar \cite{L93,LR93,LJR93,L94} is
based on the approach of KAM theory.
This method enabled numerical treatment of long-term planetary evolution in terms of a perturbed Hamiltonian system using the idea that, if a quasi-periodic function is given numerically on the complex domain, then it is possible to approximate it via a quasi-periodic function with an accuracy higher than that given by standard Fourier series.     
Namely, the quasi-periodic function is represented over a finite time interval as a finite number of terms \cite{L93}
\begin{equation}
z_j(t)=z_0 e^{i\nu_j t} + \sum_{k=1}^{N} a_{m_k} e^{i<m_k, \nu > t}
\,.
\end{equation}
Then the frequencies and complex amplitudes are computed via an iterative procedure. For example, the first frequency is determined by the maximum amplitude of $\phi(\sigma)$
\begin{equation}
\phi(\sigma) = <1/2\pi \int_{-T}^{T} f(t) e^{i\sigma t} \chi(t) dt>
\,,
\end{equation}
where $\chi(t)$ is an even-weight function.

\begin{figure}[htp]
\begin{center}
\includegraphics[height=8cm,width=12cm,angle=0]{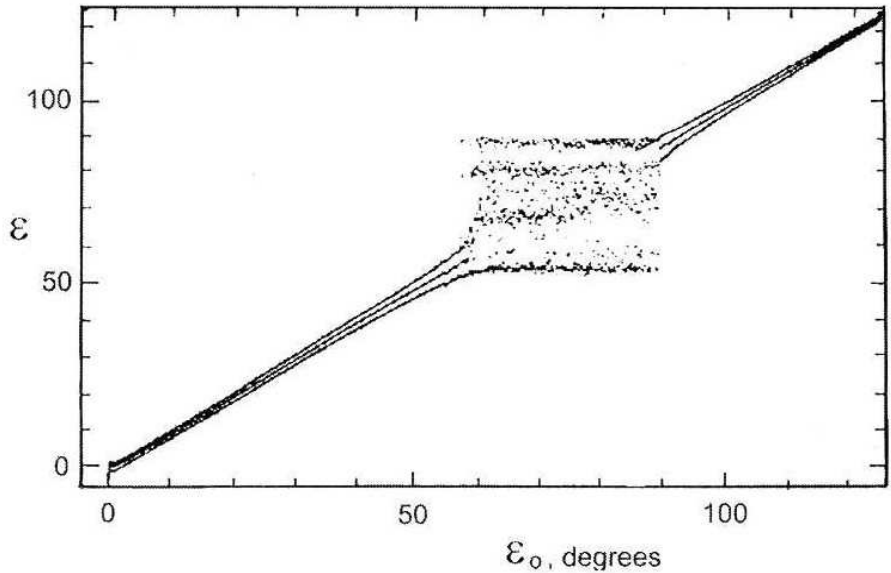}
\caption{The maximum, mean and minimum obliquity variations of the Earth depending on the initial obliquity $\epsilon_0$. Note the chaotic zone at 
$\epsilon_0=60^{\circ} - 90^{\circ}$ and the stability for its other values. The present obliquity of the Earth is within the stability zone due to the presence of the Moon, and transfers into the chaotic zone at the absence
of the Moon (Laskar et al 1993).}  
\end{center}
\end{figure}

\begin{figure}[htp]
\begin{center}
\includegraphics[height=8cm,width=12cm,angle=0]{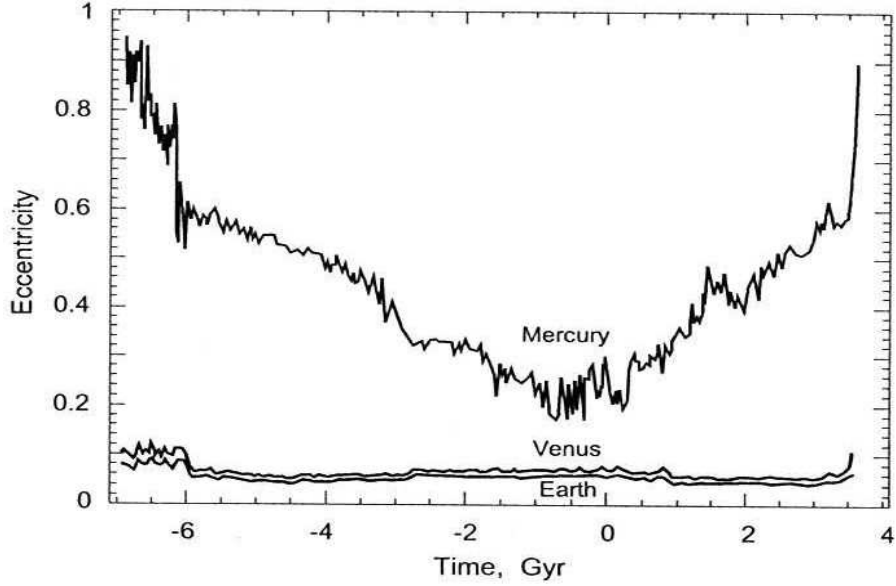}
\caption{The variation of the eccentricities of Mercury, Venus and Earth by time. The chaotic variations are suppressed for the Earth and Venus, but are significant enough for Mercury, and will lead to its escape from the Solar system within a period less than 3.5 Gyr (Laskar 1994).}
\end{center}
\end{figure}

\begin{figure}[htp]
\begin{center}
\includegraphics[height=8cm,width=12cm,angle=0]{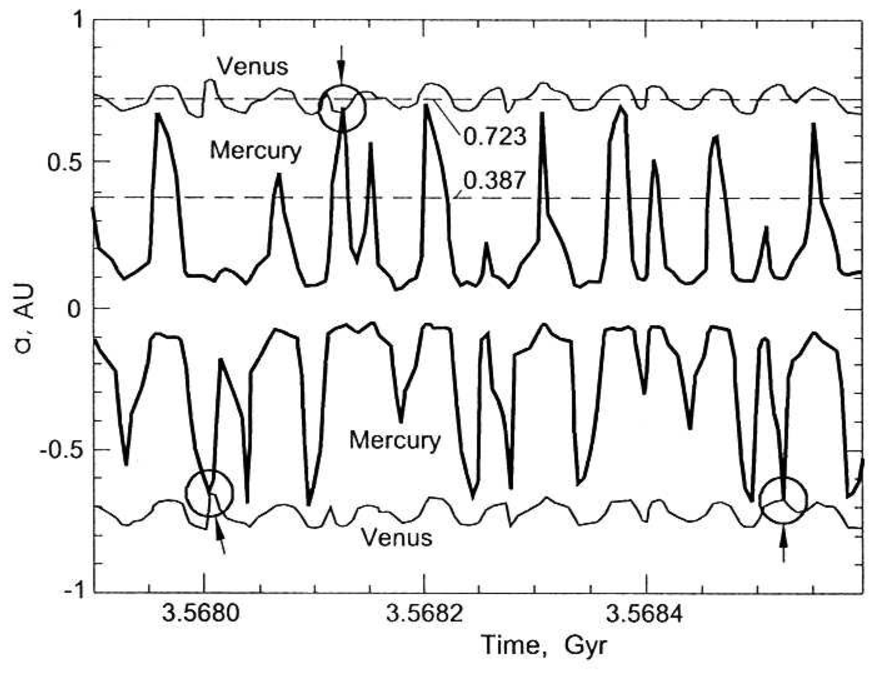}
\caption{The overlapping of the major semi-axes of orbits of Mercury and Venus due to chaotic effects (Laskar 1994).}
\end{center}
\end{figure}


Numerical integration with a time step of 500 years over the time span of about 200 million years reveals that the inner planets of the Solar system are chaotic, due to the presence of two secular resonances, one due to Mars and Earth at $\theta = 2(g_4-g_3) - (s_4 - s_3)$ 
\typeout{ did you mean sigma? since there is no theta here}
and another due to Mercury, Venus and Jupiter at $\sigma = (g_1 - g_5) - (s_1 - s_2)$, where $g_i$ and $s_i$ are the frequencies of the perihelions
and nodes, respectively. 

Laskar's calculations revealed {\it the chaotic behavior of Mercury's orbit with eccentricity variations up to 0.05, which results in its overlapping with the orbit of Venus and with inevitable escape of Mercury from its orbit.}  Chaotic behavior was discovered also for the obliquities of the planets, particularly for Mars, varying from 0 to 60 degrees. Obviously, this fact has to be taken into account while studying the past evolution of the atmosphere of Mars. {\it Chaotic behavior of the obliquity of the Earth would range even wider, within 0 and 85 degrees, with all dramatic consequences for the climate of the Earth, however, only at the absence of the Moon.
The Moon therefore, is damping the obliquity variations up to 1.3 degrees, thus stabilizing the Earth's 
climate\cite{LJR93}}.

We see that, the chaotic effects are not only able to influence essentially the dynamics of the Solar system but even 
the Earth's climate.

\section{Galactic Dynamics}

\subsection{N-body gravitating systems and geodesic flows}
 
Many properties of statistical mechanics of globular clusters and galaxies can be studied
considering a $N$-body gravitating system described by Hamiltonian
\begin{eqnarray}
\label{ham}
  H(p,r) &=& \sum_{a=1}^{N} \sum_{i=1}^{3} \frac{p^2_{(a,i)}}{2m_a} + U(r),\\
  U(r) &=& -\sum_{a<b}\frac{Gm_am_b}{|r_{ab}|},\\
  & & \quad r_{ab}=r_a-r_b.
\end{eqnarray}
We will use a well known method existing in classical mechanics,  the
Maupertuis  principle\cite{arn},   enabling   one   to   represent
a Hamiltonian system as a geodesic flow  on  some  Riemannian
manifold. In physical problems this 
approach was firstly used by Krylov\cite{Krylov} and in stellar dynamics
in\cite{GS1,GS2}. 
 
By means of the Maupertuis principle the Hamiltonian equations
\begin{equation}
\dot{r}^{\mu}=\frac{\partial H}{\partial p_\mu},\qquad
\dot{p}_{\mu}= - \frac{\partial H}{\partial r^\mu},
\end{equation}
are reduced to the geodesic equation
\begin{equation}
\nabla_{u} u = 0
\end{equation}
on the region of configurational space
$$
M = \{W=E-V(r) > 0\}
$$
with the Riemannian metric
\begin{equation}
ds^2= [E-V(r_{1,1},...,r_{N,3})]\sum_{a=1}^{N}\sum_{i=1}^{3} (dr_{a,i})^2,
\end{equation}
where $E$ is the total  energy of the system.
The condition of conservation of the total energy of the system
\begin{equation}
H(p,q) = E
\end{equation}
is equivalent to the condition on the velocity associated with the geodesic
\begin{equation}
\parallel u\parallel =1,
\end{equation}
while the affine parameter along the geodesic is determined by
\begin{equation}
ds=\sqrt {2} (E-V(r)) dt.
\end{equation}
 
The statistical properties of the geodesic  flow
are determined from the Jacobi equation
\begin{equation}
\nabla_{u}\nabla_{u}n + Riem(n,u)u=0.
\end{equation}
For a vector field satisfying the orthogonality condition
\begin{equation}
<n,u>=0,
\end{equation}
the Jacobi equation can be written in the form
\begin{equation}
\frac {d^2\parallel n \parallel ^2}{ds^2}=
     -2K_{u,n}\parallel n \parallel ^2
    + 2\parallel \nabla_{u} n \parallel ^2,
\end{equation}
where $K_{u,n}$ is the two-dimensional curvature
\begin{equation}
K_{u,n}=\frac{<Riem(n,u)u,n>}
{\parallel u \parallel ^2\parallel n \parallel ^2 - <u,n>^2} =\\
\frac{<Riem(n,u)u,n>}
{\parallel n \parallel ^2}.
\end{equation}
 
Jacobi equation  has the solution
\begin{equation}
\parallel n(s)\parallel \geq \frac{1}{2}
\parallel n(0)\parallel \exp (\sqrt {-2k} s), \qquad s>0
\end{equation}
if 
\begin{equation}
k=\max_{u,n} \{K_{u,n}\}<0.
\end{equation}
Then, the geodesic flow is an Anosov system.

{\it Thus the negativity of the  two-dimensional  curvature  the
criterion  of the instability of the geodesic flow.}
  
\subsection{Chaos of spherical systems}
 
So, we have to calculate     the
two-dimensional curvature $K_{u,n}$ for the Hamiltonian (\cite{ham}) for the 
gravitating N-body system. The Riemann curvature has the form\cite{GS1,GS2}
\begin{eqnarray*}
Riem_{\mu\lambda\nu\rho} &=& -\frac{1}{2W}[g_{\mu\nu}W_{\lambda\rho}
 +g_{\lambda\rho}W_{\mu\nu}-
g_{\mu\rho}W_{\lambda\nu}-g_{\nu\rho}W_{\mu\lambda}]\\
& &-\frac{3}{4W^2}
[(g_{\mu\lambda}W_{\nu}- g_{\mu\nu}W_{\lambda})W_{\rho}+
(g_{\nu\rho}W_{\lambda}- g_{\lambda\rho}W_{\nu})W_{\mu}]\\
& &-\frac{1}{4W^2}
[g_{\mu\nu}g_{\lambda\rho}-g_{\mu\lambda}g_{\nu\rho}]
\parallel \partial W \parallel^2,
\end{eqnarray*}
where
\begin{eqnarray*}
g_{\mu\nu} &=& W \delta_{\mu\nu},\\
W_{\mu} &=& \frac{\partial W}{\partial r^{\mu}}, \quad
W_{\mu\nu}=\frac{\partial ^2W}{\partial r^{\mu}\partial r^{\nu}}.
\end{eqnarray*}
The analysis shows that $K_{u,n}$  is  sign-indefinite, which means
that  no universal function
$$
\tau : \hbox{(all $N$-body  systems)} \to R_+
$$
exists and hence no unique relaxation time scale can exist  for  all
$N$-body gravitating systems.
 
For spherical systems,however, the two-dimensional curvature at large $N$ limit
is shown to be strongly negative, since is determined by the scalar curvature $R$
\begin{equation}
K_{u,n}=\frac{R}{3N(3N-1)}
\end{equation}
which is negative for $N > 2$
\begin{equation}
R = - \frac {3N(3N-1)}{W^3}\left(\frac{1}{4} - \frac {1}{2N}\right)(\nabla W)^2
- (3N-1) \frac {\Delta W}{W^2} < 0.
\end{equation}
 
Based on the mixing properties of dynamical systems described in previous sections,
one can define the {\it relaxation} time for spherical systems.
The explicit formula for that relaxation time taking into account the
nonlinear interaction of all $N$ bodies of the system has been estimated in\cite{GS1,GS2} (see also\cite{lang}).
For real stellar systems its value is shorter than the two-body relaxation time scale, 
but longer than the dynamical (crossing) time.
  
The disk gravitating systems can be studied using the  Lie  algebra  of  all  vector  fields   with   zero
divergence on the two-dimensional torus $T^2$\cite{arn67}, since the kinetic
energy of the element of the moving fluid induces a  right--invariant
Riemannian metric on $SDiff(T^2)$.  The  principle  of  least  action, which
determines the  motion  of  an incompressible  fluid  in terms of the
geodesics  of this metric, plays the role  of the Maupertuis principle.
One can show that, though the motion in disk galaxies is exponentially instable, the velocity
field remains constant, so one cannot speak about a relaxation in the same
sense as for spherical systems\cite{GK88}.

\subsection{Relative chaos is stellar systems}
 
The approach described above enables to consider instability of
various configurations of stellar systems. 
Average the Jacobi equation over the
geodesic deviation vector
\begin{equation}
\frac{d^2z}{ds^2}=\frac{1}{3N}r_u(s)+<\parallel\nabla_un\parallel^2>,
\end{equation}
where
$$
n=z\hat{n}, \qquad \parallel \hat{n}\parallel^2=1,
$$
and $r_u(s)$ denotes the Ricci  curvature  in  the  direction  of  the
velocity of the geodesic ({\it Ric} is the Ricci tensor)
\begin{equation}
r_u(s)= \frac{{\it Ric}(u,u)}{u^2}=\sum_{\mu=1}^{3N-1}
K_{n_{\mu},u}(s),\qquad (n_{\mu}\bot n_{\nu}, n_{\mu}\bot u).
\end{equation}
The Ricci tensor has the expression
\begin{eqnarray*}
Ric_{\lambda\rho} &=& -\frac{1}{2W}
[\Delta W g_{\lambda\rho}+(3N-2)W_{\lambda\rho}]\\
& & + \frac{3}{4W^2}
[(3N-2)W_{\lambda}W_{\rho}]-
[\frac{3}{4W^2}-\frac{(3N-1)}{4W^2}]g_{\lambda\rho}
\parallel  dW \parallel^2.
\end{eqnarray*}
Then the criterion of relative instability is\cite{GK87}:
 
\noindent
{\it the more unstable of two systems is the one with smaller negative}
$r$
\begin{equation}
r = \frac{1}{3N}\inf_{0\leq s\leq s_*}[r_u(s)],
\qquad r < 0
\end{equation}
{\it within a given interval} $0\leq s\leq s_*$, {\it i.e., this system
should be unstable with a higher probability in the same interval}.
 
Numerical exploitation of the Ricci criterion of relative instability  for
the different models of stellar  systems  has shown that, e.g.,  a spherical
system with a central mass is more unstable than  a homogeneous  one,
spherical systems are more instable than disk-like ones, etc \cite{GK87,elzant,ezg,bek}. 

\section{Galaxy Clusters: Substructure and Bulk Flows}

N-body gravitational systems, as we saw above, are exponentially instable
systems. This fact gives a key to the possibility of reconstruction of 
certain properties of based on the limited observational information,
which usually includes the 2D coordinates and 1D (line-of-sight) velocities
and the magnitudes of the galaxies. We will show how one can 
reconstruct the hierarchical substructure and the bulk regular flows of the subgroups
in the clusters of galaxies\cite{book,BekMelk}.

The developed S-tree technique is based on the geometrical methods of theory of dynamical systems
discussed in the previous sections, namely, on the introduction of the concept
{\it degree of boundness} of N particles.

Consider two particles, so that $x_1(t)$ and $x_2(t),\  t \in (-T,T)$ are their trajectories 
when their interaction is taken into  account,
and $y_1(t)$ and $y_2(t)$,
when the interaction  is  ``switched off".  

It is easy to see that the deviation of trajectories within certain time interval
\begin{equation}
    m=\max_{i=1,2}{\cal M}(x_i(\cdot)-y_i(\cdot)),
\end{equation}
can be taken as a measure of the degree of boundness
with respect to a local norm 
\begin{equation}
    {\cal M}(x(\cdot))=\sup_{t\in (-T,T)}\{|x(t)|,|\dot{x}(t)|\}.
\end{equation}

Consider balls of radius $r$ at each point of trajectories of the two
interacting particles $x_i$. The union of those balls 
$$
    {\cal C}_i(r)=\bigcup_{t\in (-T,T)} B_{x_i(t)}(r),\ i=1,2\ .
$$
of such minimal radius $m$ which contains all
trajectories of the particles will denote the free 
corresponding particles.

Two particles are considered to be $\rho$-bound for $\rho>0$ if 
$m\leq\rho$.

This  is easily generalized  to any finite number of particles.
$N$ 
particles labeled by the  set of integers
${\cal A}=\{1,\ldots,N\}$
form  a $\rho$-bound cluster if the 
distance between the 
corresponding trajectories of the  system  of interacting particles and  free  
ones  is  less  than  the  maximal deviation of all of the particles:
\begin{equation}
    m=\max_{a \in {\cal A}}{\cal N}(x_a(\cdot)-y_a(\cdot))\leq\rho\ .
\end{equation}
              
One can then define the boundness function
so that for the given local norm ${\cal N}$
\begin{equation}
    {\cal P}_Z(Y)=\max_{a \in Y}{\cal N}(z_a(\cdot)-y_a(\cdot)),
\end{equation}
where $z_a(t), y_a(t)$ are the solutions of the systems of equations
{\bf I} and {\bf II} respectively for some time interval $(-T,T)$.
In other words the boundness of ${\cal Y}$ in ${\cal Z}$ is the maximum 
deviation of the trajectories of its particles taking into account only
internal interactions compared to the situation when interactions with
particles in ${\cal Z}$ are also included.
Our goal is to split ${\cal A}$ into
$\rho$--subsystems,
i.e., to obtain the map $\Sigma$ for this choice of
boundness function.

The definition of a  $\rho$--bound cluster
  given  above  can  be reformulated now as a set of  
corresponding  particles  $A$  being  a 
connected subgraph of the graph $\Gamma$ (equivalent the matrix)  so  that 
there is no other connected subgraph $B$ including $A:\  A \subset B.$
     If one defines {\cal P} as follows
\begin{equation}
    {\cal P}_XY=\max\limits_{\stackrel{y \in Y}{z \in X\backslash Y}}
                 \{D_{yz}\}\ ,
\end{equation}
the problem of the search of a $\rho$-bound cluster is reduced  to  that 
of a connected $\Gamma$-graph.

 The algorithm of the construction of tree-diagram based on the estimation of the
two-dimensional curvature as containing information both on the
coordinates and velocities of the all particles, is developed 
and applied to various clusters of galaxies. As a result
subgroups of galaxies, "galaxy associations", of specific dynamical properties,  are detected
in the studied Abell clusters\cite{GM}, triggering later observations
by the provided lists of galaxies (see, e.g.\cite{mario}). 
The S-tree method, together with certain general assumptions on the
velocity distribution function of galaxies can be used for the
determination of the bulk velocities of the subgroups.
     
Let us stress again, that these methods arise due to the nonlinearity of the N-body
gravitating systems.

\section{General Relativity and Cosmology}

The impossibility of the direct application of results of theory of dynamical
systems developed for Riemannian spaces 
is the main difficulty arising  while studying the chaos in 
General Relativity and cosmology where one deals with pseudo-Riemannian spaces. 
Therefore, first, one has
to reformulate the concepts described in previous sections and
applied for astrophysical Newtonian systems, for the case of pseudo-Riemannian
spaces. We will give the reformulation of the property
of hyperbolicity and the covariant definition of the Lyapunov exponents given in\cite{EFI},
which are basic concepts for the study of chaos, 
and then, using it, will consider the
stability of cosmological solutions, particularly, of inflationary ones.   

I will not discuss the mixmaster models which had essentially
provoked the studies on chaos in cosmology, since they are covered in Kirillov's lectures
at VIII Brazilian School of Cosmology and Gravitation\cite{kiril}.
For the further progress on those models in the context of Non-Abelian gauge, 
string theories and pre-Big-Bang scenarios I will refer to reviews\cite{belinski,matin,damour}.
 
\subsection{Hyperbolicity in pseudo-Riemannian spaces}

Consider a geodesic flow on $M$, i.e. a group of mappings   
$\{S^t\}$ of a space $T^{\lambda}M$
$$
T^{\lambda}M=\{(x,u); x\in T_x M, g(u,u)=||u||^2=\lambda\},\, \lambda=0,\pm1.
$$
Each mapping performs a shift of a linear element   $\xi=(x,u)$
along the geodesic on distance  $t$.
 
Let $\gamma(t)$ be a geodesic on $M$ passing by a point $x\in M$,
and $\{E_a\}$ is a fixed n-dimensional basis on $T_x M$.

Transferring  $\{E_a\}$   parallel along $\gamma(t)$, i.e. getting a basis at
every $t$, one has a Fermi basis on $T_x M$.

Each vector  $X\in T_{\gamma} M$   can be represented via Fermi basis
$$
X(t)=X^a(t)E_a.
$$
with the E-norm
$$
||X||^2_E=\Sigma (X_a)^2
$$
for basis  $\{E_a\}$.

Let $\{E_{a'}\}$ be another basis.  Then a non-singular  matrix  $\Phi_a^{b'}$ exists, 
such that
$$
E_a=\Sigma \Phi_a^{b'}E_{b'}.
$$
Since both $\{E_a\}$ and $\{E_{a'}\}$  are Fermi bases, the latter relation has
to be satisfied also for constant  $\Phi_a^{b'}$.             

Then
$$
X^{b'}(t)=\Sigma \Phi_a^{b'} X^a(t).
$$

In view of non-singularity of  $\Phi_a^{b'}$, we can write
$$
C\Sigma (X_a)^2 \leq \Sigma (X_{b'})^2 \leq C^{-1}\Sigma (X_a)^2 
$$
or
$$
C||X||_E^2 \leq ||X||_E^2 \leq C^{-1}||X||_E^2,
$$
where $C$ is a positive constant.

{\it Definition of hyperbolicity}. 
Geodesic $\gamma_x(t)=S^t(\xi), ||\dot{\gamma}_x(t)||^2=\lambda$ is   
$\lambda$-hyperbolic, if there exist subspaces
$W^s(S^t(\xi))$ and $W^u(S^t(\xi))$ and $W^0(S^t(\xi))$
of the tangent space $T_{S^t(\xi)}T^{\lambda}M$
and numbers $A 0, 0<\mu<1$, such that
$$
T_{S^t(\xi)}T^{\lambda}M=W^s(S^t(\xi)) W^u(S^t(\xi)) W^0(S^t(\xi)),
$$
$$
dS^{\tau}W^s(S^t(\xi))=W^s(S^{t+\tau}(\xi)), dS^{\tau}W^u(S^t(\xi))=W^u(S^t(\xi)),
$$
where $W^0(S^t(\xi))$  is a 1D space defined by the flow vector.
 
For each $t, \tau>0$   and for a certain basis    $\{E_a\}$ we have
$$
||dS^{\tau}v||^2_E \leq A^2 \mu^{2\tau}||v||^2_E, v\in W^s(S^t(\xi)),
$$
$$
||dS^{\tau}v||^2_E \geq A^{-2} \mu^{-2\tau}||v||^2_E, v\in W^u(S^t(\xi)),
$$
where  
$$
||v||^2_E=||d\pi_{\lambda}v||^2_E+ ||K v||^2_E,
v\in TT^{\lambda}M, \pi_{\lambda} : TT^{\lambda}M \rightarrow TM,
$$
and $K$ is the mapping of connection $\nabla$.

The definition is $\{E_a\}$ invariant.

{\it Definition} Geodesic flow is $\lambda-$hyperbolic if its all 
geodesics are $\lambda$-hyperbolic.

Jacobi field is defined along the geodesic determined by the Jacobi equation
$$
\nabla_u\nabla_u Y+ R(u,v)Y=0.
$$
Correspond now to each vector $v\in TT^{\lambda}M$ a solution $Y(t)$ 
of Jacobi equation  with initial conditions               
$$
Y_v(0)=d\pi v, \nabla_uY_v(0)=k v.
$$
The resulting mapping
$$
f  :  v \rightarrow Y_v(t)
$$
is an isomorphism and
$$
d\pi dS^t(v)=Y_v(t),
KdS^tv=\nabla_vY_v(t).
$$
From the Jacobi equation we have
$$
||dS^tv||^2_E=||Y_v(t)||^2_E+||\nabla_uY_u(t)||^2_E.
$$
the latter equation and the Jacobi one enable to check the hyperbolicity condition.

{\it Definition.} 
Lyapunov characteristic exponent for maximal geodesic $\gamma$ and vector    
$v$ is defined as 
$$
\chi(\gamma,v)=\lim_{t \rightarrow \infty} \sup\frac{ln||dS^t_{\dot{\gamma}}v||^2_E}{2t}.
$$

{\it Definition.} Geodesic  $\gamma$ is stable if for any $\epsilon>0, \exists \delta(\epsilon>0$
such that from $||v||^2_E<\delta$ follows the condition $||dS^t_{\dot{\gamma}}v||^2<\epsilon$
for any $t$.  
Otherwise $\gamma$ is unstable.     
The latter two  definitions are also basis-invariant.  

Let us now define a convenient basis. 

For arbitrary geodesic $\gamma(t)$ we choose the following orthonormal basis at point
$\gamma(t)$
$$
E_0=\gamma(0)=u, E_1,...,E_{n-1};
$$
$$
g(E_a,E_b)=
$$
where $E_a$ is a dual basis.

If the following conditions are satisfied
$$
\nabla_uE_a=0=\nabla_uE^b,
$$
then the basis on $T^0M$ can be defined as
$$
E_0=u, g(E_a,E_b)=
$$
and on $T^1M$
$$
E_0=u, g(E_a,E_b)=
$$
For the vector field
$$
Y(t)=Z^a(t)E_a
$$
the Jacobi equation can be written in the form
$$
Z^a(t)+K^a_b(t)Z^b(t)=0,
$$
where
$$
K^a_b=<E^a, R(u,E{-b})u>=R^a_{bcd} u^cu^d.
$$
For the above defined basis on $T^0M$ and the Jacobi equation we have
$$
\dot{Z}=1.
$$
which means that none of geodesic flows can be 0-hyperbolic.

It can be shown that the definitions given above for spaces
of Lorentzian signature (-,+,...,+), can be generalized for  
the signatures (-,…,-, +,…,+).

The covariant definition of hyperbolicity and Lyapunov exponents given above 
enable the consideration of the stability problem of cosmological solutions.

\subsection{The ADM principle and geodesic flows in Wheeler-DeWitt superspace}

The problem of the stability of cosmological solutions is a problem of stability
in Wheeler-DeWitt superspace. We will consider this problem using the 
method of geodesic flows\cite{ADM}. 
So, first, we have to define the Hamiltonian system,
then reduce it to a flow of geodesics using the definition of the hyperbolicity given
in the previous section. 

Arnowitt-Deser-Misner (ADM) method provides the scheme of the sought Hamiltonian formulation,
assuming as given the 3-geometries of the initial and final Cauchy hypersurfaces.
We will consider locally isotropic and homogeneous cosmological models with scalar 
field when the metric can be given as
$$
h_{ij}=\sigma^2\bar{h}_{ij}
$$
where
$$
\sigma^2=\frac{4\pi G}{3}[\int \bar{h}^{1/2}d^3x]^{-1}, \, \bar{h}=det \bar{h}_{ij}
$$
We consider the Lagrangian of the scalar field
$$
L_{\phi}=-g^{-1/2}[\phi_{,a}\phi_{,b}g^{ab} + V(\phi)]
$$
and the action
$$
I=\int p_{\alpha}d\alpha + p_{\chi}d\chi-N \rm{H}_{ADM} dt
$$
where the ADM Hamiltonian is
$$
\rm{H}_{ADM}=\frac{1}{2} e^{-3\alpha}[-p^2_{\alpha} +p^2_{\chi}]+e^{3\alpha}[U(x) -\frac{k}{2}e^{-2\alpha}],
$$
where
$$
\alpha=ln a, \, \chi=\sigma \phi, U(x)=\frac{4\pi G \sigma^2}{3}V(\phi).
$$
As usual variation with respect the lapse function $N$ leads to the condition
$$
\rm{H}_{ADM}=0.
$$
To reduce the Hamiltonian system to the geodesic flow let us split the hypersurface into 
the following regions
$$
W^+\{x|V(x)>0\},\, W^-\{x|V(x)<0\},
$$
so that if the metric in region $W^-$ is Riemannian, then
$$
g_{ab}v^av^b=-2V>0, d\tau=(g_{ab}u^au^b/-2V)^{-1/2}ds
$$
and we can write the variation
$$
extI|_{H=0}=ext\int(-2V)^{1/2}(g_{ab}u^au^b)^{1/2}ds=ext\int(G_{ab}u^au^b)^{1/2}ds.
$$
Here $G_{ab}=-Vg_{ab}$ is also a Riemannian metric.

Choosing the affine parameter $s$ in order to satisfy the
condition
$$
||u||^2=G_{ab}u^au^b=1
$$
we have
$$
G_{ab}u^au^b=-Vg_{ab}v^av^b(d\tau/ds)^2=2V^2(d\tau/ds)^2.
$$
Reparameterizing the affine parameter 
$$
ds=2^{1/2}(-V)d\tau,
$$
we arrive at the flow of geodesics in the region $W^-$
$$
\rm{H}=1/2g^{ab}p_ap_b+V\rightarrow\{G_{ab}=-Vg_{ab},\, ds=2^{1/2}(-V)d\tau, ||u||^2=1\}.
$$
As regards for the region $W^+$, the classical system cannot end up in it.

Now, if the metric $g$ is pseudo-Riemannian, following the same scheme we end up
with the geodesic flow
$$
\rm{H}\rightarrow \{|V|g_{ab},\, 2^{1/2}|V|d\tau, -signV\}.
$$
Thus, we reduced the ADM Hamiltonian system to a geodesic flow on a pseudo-Riemannian manifold.

To study the stability of the geodesic flow we have to proceed from the Jacobi equation,
which has the form
$$
\frac{d^2z^i}{d\tau^2}+\gamma(z) \frac{dz^i}{d\tau}+\omega^i_j(z)=0,
$$
where
$$
\gamma=-\frac{d}{d\tau}ln|V|,
\omega^i_j=2V^2K_j^i, i,j,=1,2...k-1.
$$
Using the variables
$$
z^i=AY^i
$$
we arrive to the equation
$$
\dot{Y}^i+ 2(\dot{A}/A+\gamma)\dot{Y}^i+[(\dot{A}/A+\gamma\dot{A}/A))\delta_j^i + \omega^i_j]Y^j=0
$$
where 
$$
\dot{A}/A=-1/2\gamma, A=|V|^{1/2}.
$$
Particularly for the case $k=0$ we have the simplified expressions
$$
G=e^{6\alpha}|U|, |V|=e^{3\alpha}|U|,
$$
$$
\gamma(\tau)=-3\frac{d\alpha}{d\tau}-\frac{1}{U}\frac{dU}{d\tau}
$$
$$
K(\tau)=-\frac{1}{2} \frac{D ln G}{G}=-\frac{1}{2}\frac{ln D |V|}{e^{6\alpha}|U|},
$$
$$
\omega(\tau)=UD ln|U|, D=-\frac{\partial^2}{\partial\alpha^2}+ \frac{\partial^2}{\partial\chi^2}.
$$

\subsection{Stability of inflationary solutions}

Consider the scalar field
$$
U=\lambda \chi^n/n
$$
and the conditions
$$
\dot{H}<<H^2, \dot{\chi}^2<<U, |\dot{\chi}/\chi|<<H
$$
and $\chi>>1$ at $\alpha\rightarrow\infty$.
Then we have
$$
\dot{\gamma}=-3\dot{H}, \frac{1}{2}\dot{\gamma}+\frac{\gamma^2}{4}\simeq\frac{9}{4}H^2,
$$
and the Jacobi equation
$$
\dot{Y}^i-[-\frac{nU}{\chi^2}+(\frac{3}{2}H)^2]Y=0.
$$
From the Einstein equation
$$
\dot{H}+3H^2=6U,
$$
and in view of the condition $\dot{H}<<H^2$ we have $H^2\simeq 2U$.
If $\chi>>(2n/3)^{1/2}$, then $nU/\chi^2<<9/4H^2$, and
we have the Jacobi equation in a simple form
$$
\dot{Y}-(9/4\dot{\alpha}^2)Y=0.
$$
From its solution we obtain
$$
\delta^2=z^2+\dot{z}^2\simeq const\, U + const\, \frac{U^{'4}}{U^2}\simeq const\, \chi^n+const\,\chi^{2n-4}.
$$
We see that  $\delta$ decreases for any $n>2$ and therefore we have Lyapunov stability of the
inflationary solutions. The last formula enables also to obtain the law of the decay of perturbations at various
$n$. For example at $n=4$ we have exponential decay of perturbations, and the larger is $n$, the more stable
the solution is.

Thus we showed how one can deal with the stability problem in pseudo-Riemannian spaces,
and illustrated this on inflationary solutions.

\section{Instability in Superspace}

How typical is the given cosmological solution? This is a basic question posed since the early days of the study of the Einstein equations. In the context of later developments, particularly in quantum cosmology, the question can be reformulated in the form: to what degree are the minisuperspace models typical in superspace given the huge extrapolations involved? The consideration of perturbed minisuperspace models by Hawking and other authors still involves  extrapolation in the absence of a deeper merging of quantum theory and gravity.

The study of dynamics in the Wheeler-DeWitt superspace, more precisely, the properties of geodesic flows in superspace can provide a more general view of how typical the minisuperspace models are. 
The problem, however, is far more difficult than the one posed in conventional hyperbolicity theory, since one deals both with infinite dimensional and pseudo-Riemannian manifolds.

However, hyperbolicity  can be defined for such manifolds and we will consider the case of homogeneous cosmological models \cite{WDW}. 

The metric of Wheeler-DeWitt superspace is
$$
G^{ijkl}=\frac{1}{2}[\gamma^{ik}\gamma^{jl}+\gamma^{il}\gamma^{jk}-2\gamma^{ij}\gamma^{kl}],
$$
where
$$
\gamma= det\gamma_{ij}, i,j,k,l=1,...,n.
$$
Then one can derive
$$
G^{ijkl}d\gamma_{ij}d\gamma_{jl}=-d\xi^2 + \frac{n\xi^2}{16(n-1)}
tr (\gamma^{-1} \partial\gamma /\partial\xi^A \gamma^{-1} \partial\gamma / \partial\xi^{B})\, d\xi^A d\xi^{B}\,
$$
where
$$
\xi = 4((n-1)/n)^{1/2} \gamma^{1/4}\,,\ A, B = 1, ..., n(n+1)/2 -1\,,
$$
Then, moving to the subspace of superspace
$$
\bar{W} = (\gamma_{ij}; \gamma_{ij} \in W, \gamma = 1)
$$
with a metric induced by the metric of the superspace $W$ 
$$
G_{AB} d \xi^A d\xi^B = tr (\gamma^{-1} \partial\gamma /\partial\xi^A \gamma^{-1} \partial\gamma / \partial\xi^{B})\, d\xi^A d\xi^{B}\,,
$$
the existence of non-zero Lyapunov numbers
$$
\Sigma_i \omega_i = - n/4 < 0\,
$$
can be shown for the solutions of  the Jacobi equation
$$
z^a_i(t) = x^a_i \exp[(\pm(-\omega_i)^{1/2}t]\,.
$$
This implies the exponential instability of the geodesic flow in that subspace of the superspace.

For models with a scalar field Armen Kocharyan\cite{Koch} 
was able to show that the instability is exponential if:
\begin{enumerate}

\item
Gravitational and matter fields vary quickly with respect 
the potential;

\item
The Universe undergoes inflation in a local domain.

\end{enumerate}

The smaller are the dimension and the number of scalar fields,
the stronger is the instability.

These results lead to the following general conclusion:

{\bf The quantized system in a finite-dimensional
submanifold is not typical to that in 
superspace due to the existence of virtual perturbations along the
frozen directions which are unstable.}

This implies that minisuperspace models cannot be considered
as fair approximations of superspace models.

\section{Conclusion}

I stop here. As we saw, chaos is an inevitable ingredient of the Universe, but
it needs particular efforts to deal with. 

The spectrum of the methods applied will obviously grow further. I conclude with
mentioning the use of Kolmogorov complexity (algorithmic information) for 
the study of properties of Cosmic Background Radiation and
relations of the latter with thermodynamical and cosmological arrows.\cite{epl,boom,arrow}

\end{document}